# INTELLIGENT DATA IN THE CONTEXT OF THE INTERNET OF THINGS

Rakhi Misuriya Gupta
Open Group Certified Distinguished Architect,
Tata Consultancy Services
New Delhi
rakhi.gupta@tcs.com

**Abstract—** Advent of the Internet-of-Things will allow us to optimize equipment and resource usage, enabling increased efficiencies in automation and enabling new and more cost efficient business model. As tremendous growth opportunities emerge, so do the challenges such as diverse devices spanning across multiple networks, the need to manage the exponential growth of sensor generated data and to make sense of the huge influx of data in meaningful ways. The multitude of diversity can best be addressed by fundamentally opening up systems, architecture and applications. To go the next step and truly exploit the value of the sensor data would further require real-time analytics to gain intelligence and respond to events as they happen. Historical analysis can be used to look for trends, analyze collections of sensor data for correlation and formulate hints and suggestions based on usage and patterns. In this paper, we present a framework that overcomes diversity through its ability to flexibly represent sensor data on the internet. Business goals-driven information processing, information derived intelligence and information control as elements of the framework can further enable creation of new and innovative applications that enhance and exploit the value of Internet-of-Things.

**Keywords—** Framework for Intelligent Data in the context of Internet of things, Smart Data, Internet of Things, oData, SBVR, SDN, OGC-SWE, SensorML, Zeroconf, MQTT, CoAP, STOMP, MSDA, Business Goals driven Internet Of things.

## Introduction

We refer to the definition formulated in the Cluster of European Research Projects on the Internet of Things (CERP-IoT 2009) and describe the essence as "Smart Things - physical or virtual entities that are capable of being identified, they react to the information by triggering actions and exchanging information among themselves, as well as with the external computing entities". In addition, we refer to the IoT Toolkit (http://iot-toolkit.com/) that defines Internet-of-Things as Smart Objects consisting of encapsulation of a set of observable and semantic properties and accessible through RESTful API. Others have elaborated further on smart objects in the context of internet of things as referenced in [13], [16], [17] and [20].

We extend these definitions further to address the complete dimensions and complexities of IoT and define the objective of IoT: *To bring about tangible business benefits and cost reductions from optimization of assets enabled by a mechanism for information to be leveraged ubiquitously from one or more sources to enable more informed and intelligent decisions at various levels of sophistication. Levels of Sophistication correspond to the goals and drivers behind IoT, and may range from simple data analysis to rules that drive events ;Events that leverage information from past and present to arrive at next best action. In some cases the sophistication levels may be for semantic and cognitive computing - using artificial intelligence to extract data instead of making only simple predictions".* The sophistication levels should be able to evolve unconstrained as long as the underlying elements are well defined and have inbuilt/inherent flexibility to capture all the areas/concerns that an IoT architecture should address. Whereas a state of the art report on "Internet-of-things market, Value Networks and Business Models" reference [12] covers almost all aspects of common IoT scenarios as well as a framework we build on this to identify what aspects need to be flexible and how the implementation itself needs to be driven through the goals of IoT.

In this paper, we examine some scenarios for IoT across industries and arrive at a set of common foundational elements. A flexible framework for intelligent data on the internet is conceptualized such that new more innovative models can be created by extending it. Wherever, standards exist we map the framework to the standards best suited to address the requirement as of this writing.

## I. INDUSTRY APPLICATIONS FOR THE INTERNET OF THINGS

In this section, we first establish the various common elements of IoT through examining scenarios.

### A. Transportation and Mobility Scenario

We describe an IoT scenario of an automated transport system, where the goals is to adjust speed as per location requirement or stop for re-fuelling based on availability (Many more scenarios for automation through IoT are also addressed in reference [1]). The abstract architecture and various elements of such a system are represented in Figure 1.

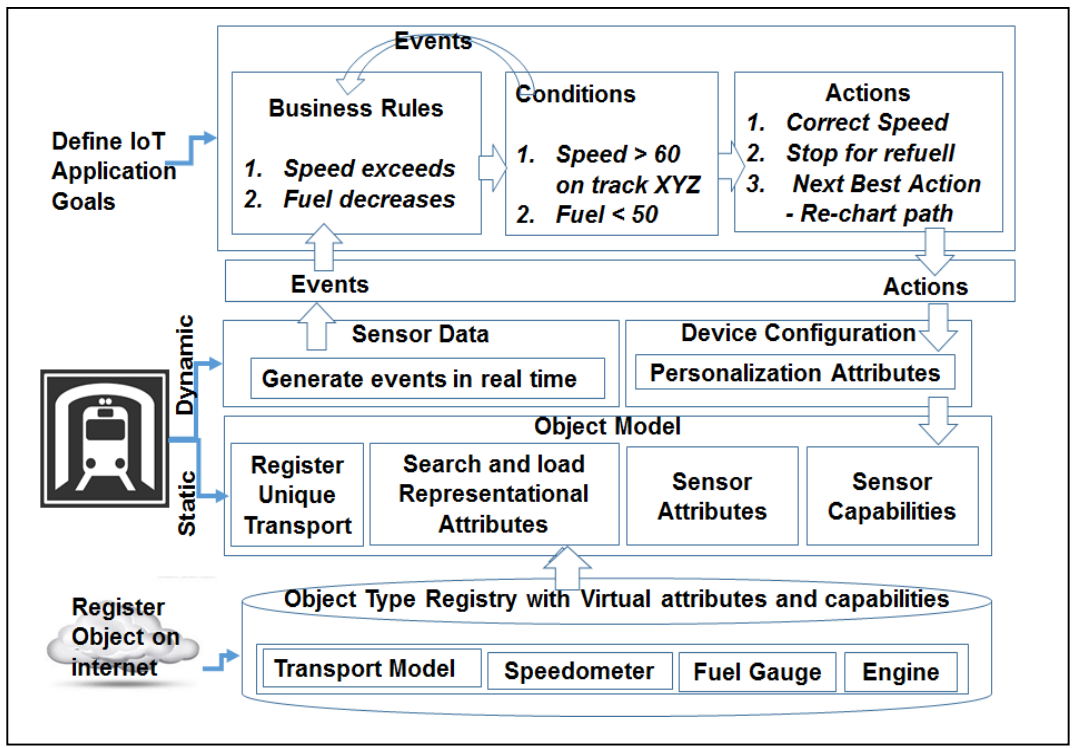

Fig. 1. Elements of an automated transport system that leverages IoT.

For the representation of the object, the EPC Network standards and infrastructure normally used in retail can be applied to track the unique automated transport object registered on the internet. When registering, additional information regarding capabilities would also be required that helps define how the data itself should be generated in episodes in case of fuel whereas continuous in case of speed or how the sensor should be controlled i.e. increase/decrease speed or halt for re-fuelling. The meta-model is best represented by the manufacturer of the transport as a registry of virtual services corresponding to the actual physical services that the different sensor of the object possesses. The configurable goals would then be to "Control speed" and "Fuel Availability", the conditions to be monitored would be speed against norms defined along different paths and fuel going below a defined threshold. Sensor data therefore uses conditions as input to scan and generate events which cause pre-determined actions to take place. If an accident takes place on a track, data feed into the system can help ensure intelligent decisions such as halting all transport on the track.

More sophisticated automation scenarios can be accomplished such as routing and increase in frequency based on data of number of passengers waiting along various tracks, accomplished through next-best action. Further sophisticated automation scenarios for wear and tear detection are possible through interpretation of sensor patterns that indicate servicing / replacement of parts obtained through machine learning and artificial intelligence.

### B. Logistics and Fulfilment Management Scenario

In logistics and fulfilment management scenario such as postal management RFID systems do tracking and tracing in real time along the mail delivery chain. A provision for context based scanning is needed in order to arrive at meaningful decisions, each scan is required to result in an event where action may need to be taken based on the category of the good such as "Express" or "Perishable". Transport optimizations maybe possible through integration with $3^{rd}$ party logistic providers. Besides meta-data and data associated with the goods, special attributes such as contextual information is required to be conveyed through the data and the associated meaning/interpretation applied through the meta-data. Service discovery and chaining become essential aspects of the framework for linkage with $3^{rd}$ party applications.

### C. mHealth Scenario in Healthcare

mHealth solutions that manage personal health through embedded systems consisting of devices with communications, processing, sensors all on a small chip. The HIPAA compliant device must collect, store or transmit sensor data. The framework must capture representational and configurable aspects of these devices as well as be able to transfer the processed data in a variety of forms either as episodes i.e. essential alerts/emergency alerts or as data press i.e. continuous processing of incoming data. A key requirement is also the need for dynamic reconfiguration of sensor networks, reference [11] deals with the issue extensively. Flexible and innovative mechanisms therefore in the mechanism by which data is processed in the IoT platform.

Another interesting application is referenced in [7].

Further the OpenIoT Consortium 2013 – www.openiot.eu conducted a requirements analysis to understand the needs for and barriers to IoT and made the results public, this can be referenced in [5].

## II. ELEMENTS OF THE FLEXIBLE NETWORK

We use the above scenarios to arrive at some common elements for the framework:

### A. Stakeholder concerns/goals

It should address the concerns/goals for which the solution was implemented. As these goals cannot be predicted in advance, we need a mechanism to flexibly define rules that govern the behavior based on events and under various conditions.

### B. Standards compliance concerns

It should provide a reference against which standards bodies can extend standards, or integrate with existing ones.

### C. Flexible Architecture

Introduces the need for an architecture that provides a uniform way of designing consistent solutions with uniformity and repeatability in the structure of the IoT solutions across the enterprise, driven through the need for predictability.

### D. Infrastructure Architecture Concerns:

Key infrastructure architecture concerns include:

- Security - Information and data Privacy, naming and identity management. data integrity, confidentiality and authentication of end-points.

- Scalability and performance - Need to support large scale deployment of devices requires a scalable platform additionally communication with devices should ensure low communication overheads. The platform must also support patterns that enable optimization of execution time. The overall scalability and performance of the platform will depend on the scalability and performance that can be achieved by each of the underlying layers. This requires device, network, storage, middleware processing and value added information processing optimizations.

- Quality-Of-Service - The different elements of the IoT must provide an ecosystem that ensure a consumer's quality of service requirements are satisfied. The risks associated with usage of ad-hoc networks can impede the Quality-of-service and hence relevant techniques are needed to counter these issues as referenced in [15].

This paper focusses on the functional requirements and the corresponding flexible framework required it does not map the above mentioned non-functional requirements of an IoT platform. We point to authoritative papers where substantial research has been carried out regarding security concerns and solutions as referenced in [3], [6], [10], [18] and [22].

### E. Data Source Support and Management

Imagine a network with zillions of devices intermittently connecting and disconnecting to the network, administration in such a situation can be very cumbersome. In addition, devices connected over area networks have varying types of resources and are unreliable, such as Nodes added / removed unpredictably. Therefore, auto-discovery of devices, ability to represent/access/control any physical device, software and sensor, embedded or tracking device on the net, both the physical as well as virtual (configurable) aspects of the device such as automated sensor configuration, as well as features such as ability to upgrade the firmware are some important aspects. Different types of sensor systems have further been elaborated in reference [21].

### F. Data Transmission Support and Management

An IoT solution may span multiple networks and may consists of groups of devices connected either directly or in an area network that in turn connects to the IoT solution. The connection between the devices / area networks are more effectively managed through gateways that help achieve the objectives of connectivity layer registration, sensor data collection, communication and protocol transformations and mappings. The core network further connects to the core IoT platform.

### G. Data Context Support and Management

The core IoT platform serves to further process the data:

- Processed Data: An IoT middleware that provides the functions of distributed interaction, composition and coordination capabilities all flexibly controlling the processing of data.

- Value added Data/Information Intelligence – Inferring new patterns by using existing information. Filtering, aggregation, interpretation of events and taking next best action based on conditions. Where filtering, aggregation etc. may again be flexibly done based on context of operation. Thus, device from same data can be used in multiple ways, depending on the configured condition and action defined.

Further discussion on context-aware sensors is referenced in [19].

### H. Domain of Participation

Defined by the entities (other devices on the internet / consumers of the information) that utilize the information generated for visualization/ making decisions/taking actions.

### III. INTERACTION AND COORELATION AMONG SOME ELEMENTS

The interaction between the "Data Source Support and Management" and "Data Context Support and Management" can be described through the development and the runtime perspectives.

- We define the development perspective to consist of a service provider/device developer who will register an object and corresponding service on the "Object/Directory Service" and "Service Registry". Additionally, a business expert will define the goals and the corresponding business description for the semantic rules which will help establish define the context in which various events will be interpreted and corresponding actions executed.

- We define the runtime perspective to consist of the business user who will request for a business service from a representational device (either he will register the device for unique identification on the web or the device is already pre-registered), and in return he will be provided by a set of capabilities/service list corresponding to the business needs.

- A Business User will request for a business service. Based on the request, a list of services and a corresponding list of capabilities that are provided by the device will be returned.

## IV. DEFINING A FLEXIBLE FRAMEWORK FOR INTELLIGENT DATA

The framework provides an abstract representation of how objects in the real world can be represented on the web. An object is, therefore, represented by a model and is associated with its corresponding ubiquitous data source (In some cases on the device, in other cases on the cloud, or elsewhere).

Additionally, the object further evolves flexibly to acquire intelligence regarding it source of transmission. The definition of the source is through Network Model. Data that gets transmitted may have been further processed / filtered to the extent as is applicable from the defined capabilities either at the level of the device or at the level of the Network. The Network Data again corresponds to a ubiquitous data source and by definition maybe simply a streaming source of real time data that is not stored or in other cases it may be stored for historical analysis.

The Device and Network Intelligent object can additionally acquire context associated intelligence. A business expert at design time specifies business rules. The Natural language definition is converted to a semantic language model to define the context in which data should be interpreted and the device and network intelligent objects now acquire contextual intelligence.

These objects can further evolve to different "Domains of Participation", text mining and analytics software can receive a large part of their knowledge from these now "Intelligence Objects on the internet". Therefore we have coined the word "Intelligent Objects" and the framework as "Framework for Intelligent Data on the internet".

A key requirement is also the need to access applications in a secure fashion and to configure various aspects of security. The above framework flexibly addresses this requirement through the ability to configure security at the level of the representational aspect, the configurability aspect and at the "Data Source", "Data Transmission" and at the "Data Context" level.

Post identification of common elements, a flexible framework for intelligent data that brings together all the elements is represented as shown in Figure 2.

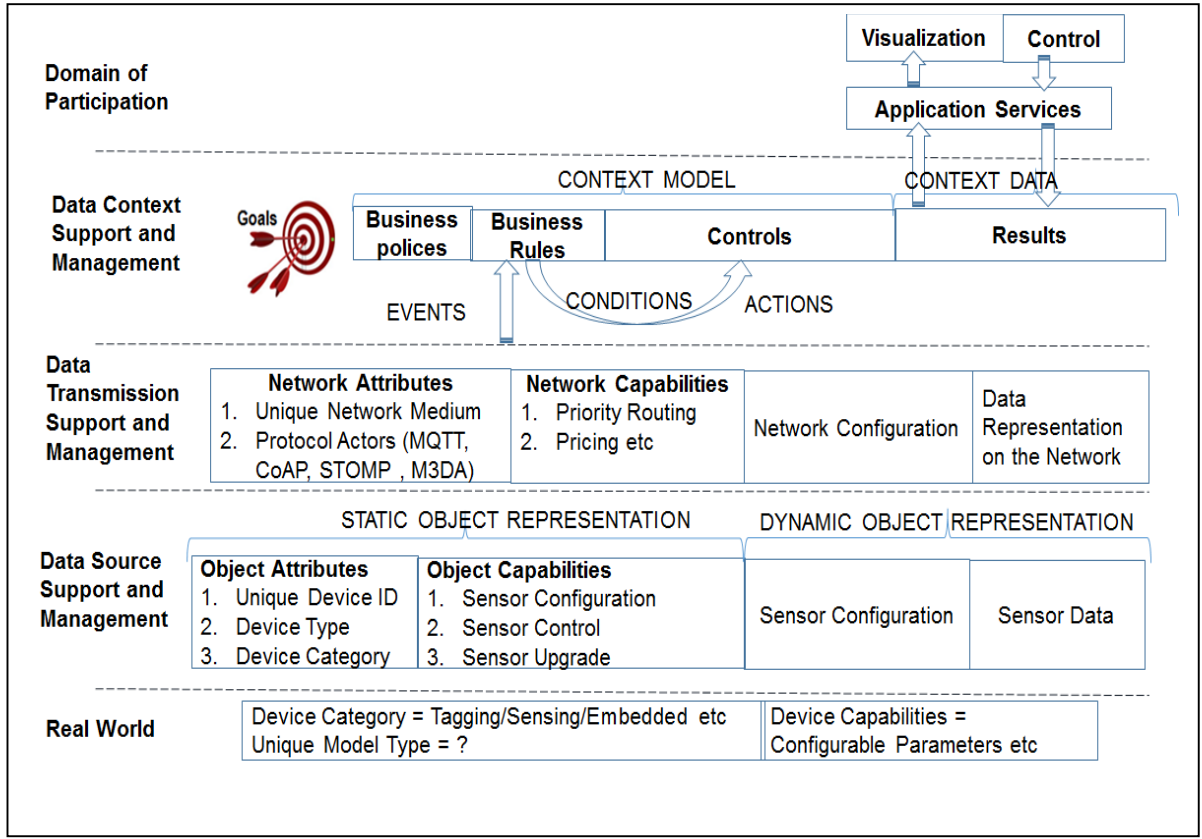

Fig. 2. A Flexible framework for Intelligent Data on the Internet

Research has been conducted around virtualization framework for real world objects and cognitive management their internet representation - reference [2]. Additionally, research has been conducted around semantic fusion based architecture for IoT - reference [14]. The key distinguishing features of the framework we have elaborated above are:

a. Business Goal driven IoT

b. Flexible control and modification of business policies and rules that allow actions based on configurable conditions

c. Layered framework with flexible extensions at each layer

d. Software driven configurations at every layer

e. Mapping to best practice standards at every layer that make the framework feasible

## V. MAPPING FRAMEWORK TO STANDARDS

We further attempt to map the flexible framework defined above to various existing standards and protocols.

### A. Data Source Support and Management

How do we enable automatic device discovery that is efficient on the network, resilient to extended periods of packet loss, simple and doesn't take a lot of code, able to function even in the face of many common network misconfigurations? Standards such as Zeroconf implement a "Service Discovery Gateway" that listens to service announcements on configured network segments and builds a cache of services and corresponding addresses. Then, it can be configured to proxy these requests to other segments. It is based on the mechanisms addressing, naming and discovery. As compared to Zeroconf, UPnP (another alternative protocol) suffers from a number of drawbacks, this discussed in reference [9].

How do we use a general and extendable mechanism that describes device capability and controls device behavior that overcomes vendor-specific interfaces and resource constraints? An important requirement is the need for flexible definition of services. Hence a resources based architecture where it is possible to expose resources and enable users to execute queries to extract information is a suitable solution. The approach is similar to database systems for extracting information using SQL.

One of the options for resources based architecture is OData (Reference [4] discusses how OData can be used to represent data on the network). OData was the initiator of using semantics to specify entities. OData is a web services compliant protocol using the WCF Data Service technology to access the device associated data it uses CRUD operations to manipulate the data. A drawback of OData is higher data and protocol complexity and therefore higher levels of energy consumption however as compared to other application level protocol it is within reasonable bounds.

A general, web oriented way and more efficient mechanism to describe a device explicitly is to use a markup language such as SensorML by OGC's Sensor Web Enablement (SWE) initiative, it uses Sensor Observation Service (SOS) for registering a service that overcomes vendor-specific interfaces and resource constraints. Another, alternative is OMA Device Management Standards (conform to TR-069 series) already widely used for managing mobile internet devices and handsets. Device provisioning, configuring, capability settings, status and information retrieving can all be achieved by OMA DM 2.0 the next generation RESTful based DM Protocol. These protocols are more energy efficient and lightweight.

### B. Data Transmission Support and Management

The mechanism for "Data Source Definition" can be extended beyond device object registration to network registration. Where different network's and the capabilities defined can be published in the "Object and Directory Services" as well as "Service Registry". The meta-data of the object can include inbuilt physical features of the transmission medium such as connectivity features etc., whereas various configurable properties such as usage and available billing models can be defined as virtual attributes that are configurable. Therefore, a transmission medium provided can register his services for usage and discovery. Other business users can discover the services and capabilities required and register their device to use the required transmission medium services to transmit their data over the web.

The association between the "Data Source" and "Data Transmission" will therefore be created as a temporary registered object in a "Network Usage Registry" and the history of the usage will be maintained along with the linkage between the "Data Context" and the "Data Source" aspects. Network can be defined further in terms of the physical aspects i.e. the unique network path to be leveraged for transmission. The virtual or capability aspects are the available translation mechanisms supported between protocols such as MQTT, CoAP, STOMP, MSDA (A survey of various technologies and protocols is covered in reference [8]). Both aspects go together to define the media that should be selected for transmission in addition to this there may be other parameters that a network vendor will configure.

Software Defined Networking (SDN) centralizes control of the network by separating the control logic to off-device computer resources, thus providing a mechanism to represent network resources. Further innovative use cases for optimizing the utilization of the network are also possible through exploiting SDN further.

### C. Data Context Support and Management

There are two aspects to this – the aspects of "Distributed Interaction, Composition and Coordination capability" and the aspects of "Context Associated Interpretation and Control". While the former is addressed through the capabilities of a middleware the later requires dynamic interpretation based on the business goals defined.

Data Semantics of Business Vocabulary and Business Rules (SBVR) is a specification defined by OMG that is suited to the requirements for data context support and management element of IoT. SBVR specifications enable business policies, business rules that governs business actions to be defined. Text mining and analytics solution can receive intelligence from SBVR defined vocabulary. SBVR concepts and concept relationships enable the ability to match them in the incoming information and produce more accurate and insightful results.

## Conclusion

An IoT platform may be deployed in multiple scenarios and use cases and requires traversing multiple stakeholders, requiring multiple protocols to be supported across the layers of the device, area networks, core networks, IoT Platform and the IoT applications. The challenges can be addressed effectively through a flexible, extendible and layered framework with minimum coupling at every layer, with physical aspects of various layers represented through virtual configurable objects as elaborated here. There are many vendors who are working on different abstraction goals and also collaborating with each other for developing the complete picture (example: BBF, ETSI M2M, HGI, OSGI, OMA), however an end-to-end framework that provides the complete picture is yet to evolve as the protocols are continuously evolving.